\documentclass[conference]{IEEEtran}
\IEEEoverridecommandlockouts

\usepackage{cite}
\usepackage{amsmath,amssymb,amsfonts}
\usepackage{graphicx}
\usepackage{textcomp}
\usepackage{xcolor}
\usepackage{booktabs}
\usepackage{multirow}
\usepackage{array}
\usepackage{balance}
\usepackage{url}
\usepackage{enumitem}
\usepackage{placeins}

\usepackage{tikz}
\usepackage{textcomp}

\begin{document}

\title{CAKE: Cloud Architecture Knowledge Evaluation of Large Language Models}

\author{
  \IEEEauthorblockN{Tim Lukas Adam}
  \IEEEauthorblockA{\textit{Centre for Industrial Software}\\
    \textit{University of Southern Denmark}\\
    Alsion 2, S{\o}nderborg, 6400, Denmark\\
  tiada23@student.sdu.dk}
  \and
  \IEEEauthorblockN{Phongsakon Mark Konrad}
  \IEEEauthorblockA{\textit{Centre for Industrial Software}\\
    \textit{University of Southern Denmark}\\
    Alsion 2, S{\o}nderborg, 6400, Denmark\\
  phkon23@student.sdu.dk}
  \and
  \IEEEauthorblockN{Riccardo Terrenzi}
  \IEEEauthorblockA{\textit{Centre for Industrial Software}\\
    \textit{University of Southern Denmark}\\
    Alsion 2, S{\o}nderborg, 6400, Denmark\\
  rite@mmmi.sdu.dk}
  \and
  \IEEEauthorblockN{Florian Girardo Lukas}
  \IEEEauthorblockA{\textit{Centre for Industrial Software}\\
    \textit{University of Southern Denmark}\\
    Alsion 2, S{\o}nderborg, 6400, Denmark\\
  fllu@mmmi.sdu.dk}
  \and
  \IEEEauthorblockN{Rahime Yilmaz}
  \IEEEauthorblockA{\textit{Centre for Industrial Software}\\
    \textit{University of Southern Denmark}\\
    Alsion 2, S{\o}nderborg, 6400, Denmark\\
  rayil@mmmi.sdu.dk}
  \and
  \IEEEauthorblockN{Krzysztof Sierszecki}
  \IEEEauthorblockA{\textit{Centre for Industrial Software}\\
    \textit{University of Southern Denmark}\\
    Alsion 2, S{\o}nderborg, 6400, Denmark\\
  krzys@mmmi.sdu.dk}
  \and
  \IEEEauthorblockN{Serkan Ayvaz}
  \IEEEauthorblockA{\textit{Centre for Industrial Software}\\
    \textit{University of Southern Denmark}\\
    Alsion 2, S{\o}nderborg, 6400, Denmark\\
  seay@mmmi.sdu.dk}
}

\maketitle

\begin{abstract}
  In today's software architecture, large language models (LLMs) serve as software architecture co-pilots. However, no benchmark currently exists to evaluate large language models' actual understanding of cloud-native software architecture. For this reason we present a benchmark called CAKE, which consists of 188 expert-validated questions covering four cognitive levels of Bloom's revised taxonomy---recall, analyze, design, and implement---and five cloud-native topics. Evaluation is conducted on 22 model configurations (0.5B--70B parameters) across four LLM families, using three-run majority voting for multiple-choice questions (MCQs) and LLM-as-a-judge scoring for free-responses (FR). Based on this evaluation, four notable findings were identified. First, MCQ accuracy plateaus above 3B parameters, with the best model reaching 99.2\%. Second, free-response scores scale steadily across all cognitive levels. Third, the two formats capture different facets of knowledge, as the MCQ accuracy approaches a ceiling while free-responses continue to differentiate models. Finally, reasoning augmentation (+think) improves free-response quality, while tool augmentation (+tool) degrades performance for small models. These results suggest that the evaluation format fundamentally shapes how we measure architectural knowledge in LLMs.

\end{abstract}


\section{Introduction}

Recently, Large language models (LLMs) are becoming co-pilots for software engineering, from writing code to making architectural decisions~\cite{b12}. Cloud-native software architecture covers microservices, containerization, orchestration, and cloud deployment patterns. In this domain, architectural choices have consequences for scalability, resilience, and maintainability~\cite{b9,b10}.

Today's benchmarks test LLMs on code generation (SWE-bench~\cite{b1}, HumanEval~\cite{b2}), code reasoning (CRUXEval~\cite{b4}), or broad knowledge (MMLU~\cite{b5}, BIG-bench~\cite{b6}). ArchCode~\cite{b3} examines architecture-related code understanding but remains at the code level rather than spanning cognitive levels of architectural knowledge. No current benchmark evaluates whether LLMs grasp the conceptual and procedural knowledge behind cloud-native architecture decisions.

Software architects increasingly use LLMs throughout the software architecture lifecycle, e.g., to gather and refine requirements, explore design alternatives and trade-offs, and draft architectural documentation and decisions~\cite{b12}. Therefore, practitioners need to understand where LLM assistance is reliable and where human judgment remains essential.
With the goal of establishing a benchmark for cloud-native software engineering tasks, we address this gap with three contributions:
\begin{enumerate}
  \item CAKE, a benchmark for cloud-native software architecture knowledge across four cognitive levels of Bloom's revised taxonomy~\cite{b13}, containing 188 expert-validated questions across five cloud-native topics.
  \item Empirical findings from testing 22 model configurations (0.5B--70B parameters) across four LLM families, including base, reasoning-enhanced (+think), and tool-augmented (+tool) variants.
  \item Public artifacts, including the benchmark dataset\footnote{\url{https://github.com/timadam03/CAKE-benchmark}}.
\end{enumerate}

The remainder of this paper is organized as follows. Section~\ref{sec:related} reviews related work and Section~\ref{sec:benchmark} introduces the benchmark CAKE. Section~\ref{sec:results} reports results, Section~\ref{sec:discussion} discusses implications and limitations, and Section~\ref{sec:conclusion} presents the conclusions.

\section{Related Work}
\label{sec:related}
Recent LLM benchmarks focus on code generation and reasoning. SWE-bench~\cite{b1} tests models on real-world GitHub issues, HumanEval~\cite{b2} targets function-level code generation, CRUXEval~\cite{b4} measures code reasoning, and ArchCode~\cite{b3} introduces architecture-aware code generation tasks. All remain at the code level, none testing the architectural knowledge behind design decisions. Broad knowledge benchmarks like MMLU~\cite{b5}, MMLU-Pro~\cite{b25}, and BIG-bench~\cite{b6} cover many domains but neglect software architecture. GPQA~\cite{b7} covers graduate-level science but not software engineering. QuArch~\cite{b19} benchmarks LLMs on computer architecture — a field distinct from software architecture — nevertheless its use of Bloom's cognitive levels inspired this benchmark's design.

Domain-specific benchmarks have adopted cognitive-level frameworks. LawBench~\cite{b21} applies Bloom's taxonomy to legal reasoning. Automated Bloom's-aligned generation~\cite{b22} demonstrates that such stratification generalizes across domains. In engineering, DesignQA~\cite{b23} tests engineering documentation comprehension. A recent systematic review of software architecture and LLMs~\cite{b24} identifies evaluation of architectural knowledge as a key open challenge.

Concept inventories~\cite{b11} and Bloom's taxonomy~\cite{b14,b15} provide frameworks for multi-level knowledge testing. Bass et al.~\cite{b9} and Richards and Ford~\cite{b10} shaped our question design. On the evaluation side, LLM-as-a-judge~\cite{b8} enables scalable open-ended scoring but raises reliability concerns including language-dependent biases~\cite{b26} and prompt sensitivity~\cite{b20}; we mitigate these with a single deterministic judge and rubric-based scoring. Our pipeline combines three-run majority voting for multiple-choice questions(MCQ) with LLM-as-a-judge for free-response (FR).

\section{The CAKE Benchmark}
\label{sec:benchmark}
\subsection{Design}

CAKE targets cloud-native software architecture: microservices, containers, orchestration, and cloud-managed services. We drew 85 core concepts from several sources, including the AWS Well-Architected Framework, microservices.io, Azure Architecture Patterns, and Kubernetes Enhancement Proposals.

\begin{figure*}[t]
  \centerline{\includegraphics[width=\textwidth]{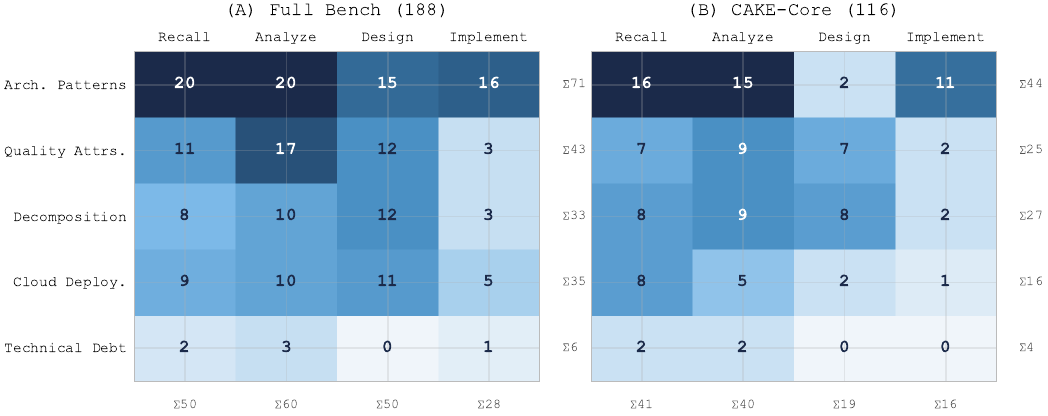}}
  \caption{Question distribution across five cloud-native topics and four cognitive levels. (A) Full Bench (188 evaluated questions; 12 implement-level MCQs excluded due to a formatting defect). (B) CAKE-Core subset (116 questions passing mean correctness $\geq$~4.0 and clarity $\geq$~4 filters, with no flags). Cell values show question counts; color intensity indicates density.}
  \label{fig:topic-skill}
\end{figure*}

Questions cover four cognitive levels mapped to Bloom's revised taxonomy~\cite{b13}: recall, analyze, design, and implement, spread across five topics: architectural patterns, quality attributes, decomposition strategies, cloud deployment, and technical debt. The benchmark holds 130 MCQ and 58 free-response questions (188 total): recall~50, analyze~60, design~50, implement~28. Implementation knowledge is assessed exclusively through free-response. The question distribution is illustrated in Fig.~\ref{fig:topic-skill}.

\begin{table}[h!]
  \caption{Evaluated Model Families and Configurations.}
  \label{tab:models}
  \centering
  \scriptsize

  \begin{tabular}{@{}c@{}}
    \begin{tabular}{llrrrr}
      \toprule
      \textbf{Family} & \textbf{Sizes} & \textbf{Base} & \textbf{+think} & \textbf{+tool} & \textbf{Total} \\
      \midrule
      Qwen    & 0.5--30B          & 6 & -- & -- & 6 \\
      Llama   & 1--70B            & 4 & 3 & 3 & 10 \\
      Mistral & 3--14B            & 3 & 1 & -- & 4 \\
      GPT     & 4o-Mini, 5-Mini   & 2 & -- & -- & 2 \\
      \midrule
      \multicolumn{2}{l}{\textbf{Total}} & \textbf{15} & \textbf{4} & \textbf{3} & \textbf{22} \\
      \bottomrule
    \end{tabular}\\[5pt] \\
    {\footnotesize \parbox{0.95\linewidth}{Count of configurations per family. Configurations consist of base (direct question answering), +think (structured reasoning with Chain-of-Thought), and +tool (agentic tool use enabled).}}
  \end{tabular}
\end{table}

\subsection{Generation and Validation}

Claude Opus 4.5 generated all questions using a parametric prompt template that specified the target topic, skill level, difficulty tier, and response format. Questions were generated in batches per topic--skill cell, with manual review between batches to check diversity.

\begin{figure}[t]
  \centerline{\includegraphics[width=\columnwidth]{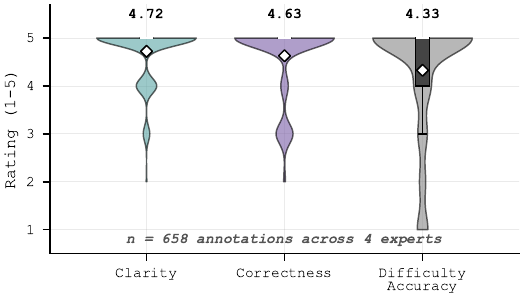}}
  \caption{Distribution of expert ratings across clarity, correctness, and difficulty accuracy. Violin plots show density; box overlays show median and IQR. High means (4.72, 4.63) with narrow spread confirm benchmark quality.}
  \label{fig:expert-ratings}
\end{figure}

Four domain experts independently rated the 200 candidate questions on clarity (1--5), correctness (1--5), and difficulty accuracy (1--5), also flagging ambiguity or typos. Three experts completed all annotations, while one provided partial ratings. In total, the 658 annotations indicate high quality (Fig.~\ref{fig:expert-ratings}): mean clarity 4.72 out of max. 5, mean correctness 4.63 out of max. 5. Pairwise agreement within one scale point reaches 91.3\% for clarity and 77.4\% for correctness. Ordinal Krippendorff's $\alpha$~\cite{b16} is near zero for both ($\alpha=0.00$ and $\alpha=-0.01$), reflecting a ceiling effect rather than genuine disagreement: when most ratings fall between 4 and~5, the remaining variance becomes difficult to interpret. After validation, we excluded 12 implementation-level MCQs due to a formatting defect in the options array (Section~\ref{sec:limitations}). Because implementation knowledge is better captured through free-response questions, 188 questions remained for evaluation.

\subsection{Models}

We tested 22 model configurations from four families spanning 0.5B to 70B parameters (Table~\ref{tab:models}). Locally-served models from the Qwen, Llama and Mistral families run on Ollama\footnote{\url{https://ollama.com}} and GPT models run on OpenRouter\footnote{\url{https://openrouter.ai}}. All use temperature~0 to reduce randomness and create more consistent outputs across runs. Three execution modes are available per model, the baseline (direct question-answer), structured reasoning (+think, turning on chain-of-thought by prepending a reasoning preamble and raising the maximum token limit from 256 to 2048), and agentic tool use (+tool, which supplies web search and browsing functions). These modes map to the config column in Table~\ref{tab:models}. The reports of the full per-configuration results are presented in Table~\ref{tab:overall}.

\subsection{Evaluation Pipeline}

The evaluation pipeline follows a provider-agnostic design supporting three inference backends: Ollama (CLI runtime), LM~Studio\footnote{\url{https://lmstudio.ai}} (GUI runtime), and OpenRouter (cloud API). Each provider module exposes a shared interface for model initialization and response generation, so models can be replaced without adjusting the evaluation logic.

The diverse output formats are extracted through a multi-stage priority chain, including explicit pattern matching, stripping of thinking tags for reasoning-mode outputs, JSON parsing, and regex matching. This layered approach is necessary because different models and configurations produce answers in varying formats.

\subsection{MCQ Evaluation}

Each MCQ is presented three times with shuffled options, with a majority voting reducing positional bias and capturing answer consistency. The score is the fraction of runs that agree with the majority answer: 1.0 (unanimous) signals high confidence, while 2/3 (split majority) flags uncertainty. Across all configurations, unanimous answers reach 89.5\% accuracy versus 55.0\% for split-majority answers, a 34.5 percentage-point gap confirming conviction as a meaningful confidence signal.

\subsection{Free-Response Evaluation}

Free-response questions are graded using LLM-as-a-judge~\cite{b8}, with DeepSeek-R1:32B as the judge model on a deterministic 0--5 grading scale. The judge receives the question, reference rubric, and model response, then produces a single numerical score. DeepSeek-R1 was chosen for its consistency and stable scoring behavior across multiple runs.
\section{Results}
\label{sec:results}

Our study results are presented in Table~\ref{tab:overall} for all 22 configurations, grouped by model family. The results are organized into four categories: MCQ performance, free-response evaluation, augmentation effects, and overall trends across formats.

\subsection{MCQ Performance}

MCQ accuracy approaches a ceiling for most configurations. Sixteen of 22 models exceed 90\%, and within each family the scaling of the base model is monotonic. Above 3B parameters, gains shrink to 1--3 percentage points per size step. GPT models achieve perfect or near-perfect MCQ accuracy. Across local models, performance converges rapidly once models exceed the 3B threshold. Resource usage varies slightly, with inference time ranging from 0.9\,s for Qwen~3B to 8.8\,s for Llama3.1~8B +tool per question. Augmented variants do not significantly increase VRAM usage, but introduce longer inference times, with +think increasing latency by approximately 3--7$\times$.

\begin{table}[t]
  \caption{Performance and resources for all 22 configurations, grouped by family ($\uparrow$~higher is better). GPT models show ``---'' for Params and VRAM as they run via OpenRouter.}
  \label{tab:overall}
  \centering
  \scriptsize
  \begin{tabular}{lccccrr}
    \toprule
    \textbf{Model} & \textbf{Params} & \textbf{Cfg} & \textbf{MCQ} & \textbf{FR} & \textbf{Time} & \textbf{VRAM} \\
    & & & \textbf{(\%)}$\uparrow$ & \textbf{Overall}$\uparrow$ & \textbf{(s)}$\downarrow$ & \textbf{(GB)} \\
    \midrule
    \multicolumn{7}{l}{\textbf{Qwen}} \\
    \cmidrule(lr){1-7}
    Qwen 2.5      & 0.5B & base   & 47.7          & 1.99 & 1.2          & 30.5 \\
    Qwen 2.5      & 1.5B & base   & 91.5          & 2.71 & 1.0          & 31.8 \\
    Qwen 2.5      & 3B   & base   & 95.4          & 3.06 & \textbf{0.9} & 28.7 \\
    Qwen 2.5      & 7B   & base   & 96.9          & 3.02 & 1.1          & 28.7 \\
    Qwen 2.5      & 14B  & base   & 97.7          & 3.36 & 1.5          & 28.7 \\
    Qwen3-Coder   & 30B  & base   & \textbf{99.2} & 4.04 & 1.8          & 29.0 \\
    \midrule
    \multicolumn{7}{l}{\textbf{Llama}} \\
    \cmidrule(lr){1-7}
    Llama 3.2     & 1B  & base    & 37.7          & 2.38 & 1.0  & 28.7 \\
    Llama 3.2     & 1B  & +think  & 76.2          & 2.76 & 4.4  & 24.2 \\
    Llama 3.2     & 1B  & +tool   & 13.8          & 1.71 & 2.5  & 24.2 \\
    Llama 3.2     & 3B  & base    & 93.8          & 2.86 & 1.3  & 28.7 \\
    Llama 3.2     & 3B  & +think  & 93.8          & 3.01 & 4.9  & 25.4 \\
    Llama 3.2     & 3B  & +tool   & 40.8          & 1.75 & 2.5  & 25.4 \\
    Llama 3       & 8B  & base    & 93.8          & 2.96 & 1.2  & 28.7 \\
    Llama 3       & 8B  & +think  & 95.4          & 2.94 & 6.3  & 28.7 \\
    Llama 3.1     & 8B  & +tool   & 91.5          & 2.85 & 8.8  & 28.1 \\
    Llama 3.3     & 70B & base    & \textbf{99.2} & 3.47 & 3.7  & --- \\
    \midrule
    \multicolumn{7}{l}{\textbf{Mistral}} \\
    \cmidrule(lr){1-7}
    Mistral 3.1   & 3B  & base    & 96.9          & 3.54 & 1.3  & 29.0 \\
    Mistral 3.1   & 3B  & +think  & 83.8          & 3.96 & 8.6  & 26.3 \\
    Mistral 3.1   & 8B  & base    & 98.5          & 4.08 & 1.7  & 29.0 \\
    Mistral 3.1   & 14B & base    & \textbf{99.2} & \textbf{4.33} & 2.4 & 29.0 \\
    \midrule
    \multicolumn{7}{l}{\textbf{GPT}} \\
    \cmidrule(lr){1-7}
    GPT-4o-Mini   & --- & base    & \textbf{99.2} & 3.77 & 1.9  & --- \\
    GPT-5-Mini    & --- & base    & \textbf{99.2} & \textbf{4.52} & 10.0 & --- \\
    \bottomrule
  \end{tabular}
\end{table}

\subsection{Free-Response Analysis}

Free-response scores span a much wider range than MCQs, from 1.71 for Llama3.2~1B +tool to 4.52 for GPT-5-Mini. This spread exposes capability gaps that MCQ results do not capture. Implement-level free-response is the primary differentiator. The score range at this level, 1.36--4.54, exceeds the spread at analyze, 2.00--4.75, and design, 1.31--4.31. Implementation is evaluated exclusively through free-response, reinforcing the value of generative evaluation for higher-order architectural tasks.

\begin{figure}[htbp]
  \centerline{\includegraphics[width=\columnwidth]{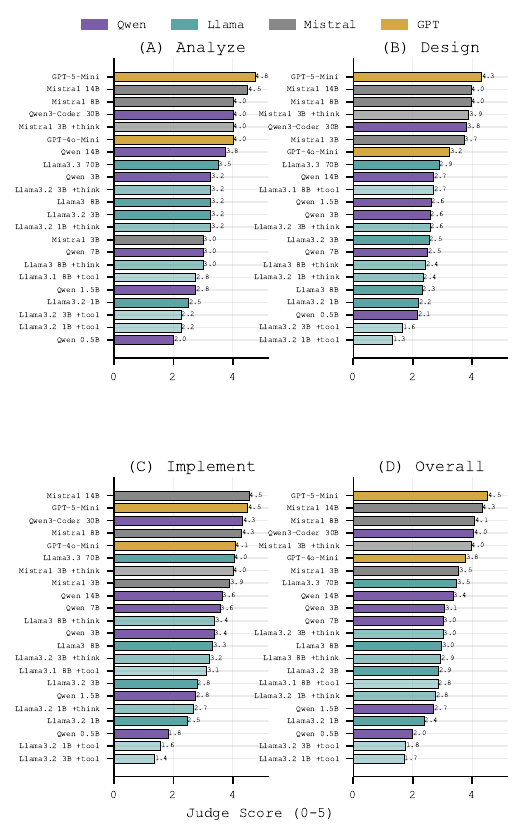}}
  \caption{Free-response judge scores (0--5) for all 22 configurations, ranked per panel, (A) Analyze, (B) Design, (C) Implement and (D) Overall. 
  \\Each panel shows one cognitive level; color indicates model family. Mistral models consistently rank highest among local models.}
  \label{fig:fr-overview}
\end{figure}

Fig.~\ref{fig:fr-overview} ranks all configurations by judge score across cognitive levels. Mistral~14B and GPT-5-Mini consistently occupy the top two positions. Rankings vary across panels, indicating that cognitive-level scores are not perfectly correlated. GPT-4o-Mini places 6th on analyze but 5th on implement, while Llama3.3~70B ranks higher on implement than on design. The gap between MCQ and free-response performance is most pronounced for mid-size models in the 3--8B range, which often select correct MCQ answers but struggle to articulate architectural reasoning.

Within each family, free-response performance increases with parameter count, though gains diminish at mid-range sizes. In the Qwen family, FR overall rises from 1.99 at 0.5B to 3.06 at 3B and 4.04 at 30B, with a slight decline between 3B and 7B ($-$0.04). Llama shows a similar plateau, increasing from 2.38 at 1B to 2.86 at 3B and 2.96 at 8B. Mistral displays steady improvements at every size tier, from 3.54 at 3B to 4.08 at 8B and 4.33 at 14B. This family-specific scaling pattern suggests that architectural knowledge gains depend on training data composition~\cite{b18} rather than parameter count alone.

\subsection{Augmentation Effects}

\begin{figure}[t]
  \centerline{\includegraphics[width=\columnwidth]{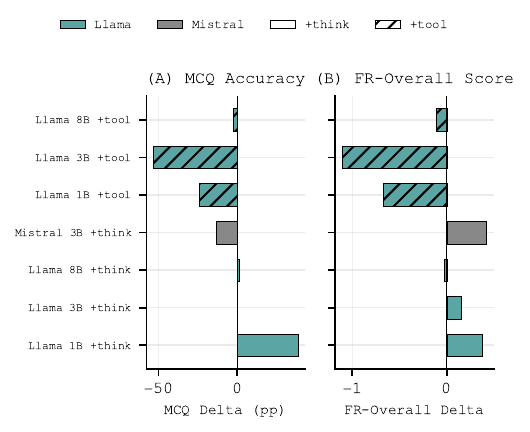}}
  \caption{Augmentation effects on MCQ and Free-response performance. 
  \\(A) MCQ accuracy delta. (B) FR overall delta for +think and +tool variants relative to their base models. Color indicates model family, hatching indicates +tool.}
  \label{fig:augmentation}
\end{figure}

MCQ and FR deltas for the four \texttt{+think} and three \texttt{+tool} variants against their base models are plotted in Fig.~\ref{fig:augmentation}.

\texttt{+think} generally improves free-response quality (+0.15 to +0.42 FR overall), with only Llama3~8B showing a marginal decline ($-$0.02), while MCQ effects are mixed. Llama3.2~1B gains +38.5\,pp while Mistral~3B drops $-$13.1\,pp. A size-dependent trade-off can be observed. For Llama3.2~1B, reasoning augmentation nearly doubles MCQ accuracy (37.7\%$\to$76.2\%) as well as lifting free-responses (+0.38), suggesting that chain-of-thought can compensate for limited parametric knowledge. At 3B, the MCQ benefit disappears (93.8\%$\to$93.8\%) while free-response still gains (+0.15). For Mistral~3B, +think degrades MCQ results by 13.1\,pp despite producing the largest free-response gain (+0.42).

\texttt{+tool} negatively impacts performance for small models. The degradation follows a clear size gradient: Llama3.2~1B drops $-$23.9\,pp MCQ and $-$0.67 FR overall, Llama3.2~3B drops $-$53.0\,pp and $-$1.11, while Llama3.1~8B loses only $-$2.3\,pp MCQ and $-$0.11 FR overall. Only the 8B variant keeps reasonable performance (91.5\% MCQ, 2.85 FR overall), pointing to a minimum capacity threshold around 8B parameters for effective tool use.

\subsection{Overall Performance}

Across formats, GPT-5-Mini achieves the highest free-response score at 4.52, while Mistral~14B leads among local models at 4.33. MCQ performance converges rapidly above 3B parameters, whereas free-response continues to differentiate models across the full size range.

Mistral achieves the strongest free-response results at each local size, with Mistral~3B outperforming larger Qwen~7B and Llama3~8B models. GPT models achieve near-ceiling MCQ accuracy yet diverge substantially on free-response, indicating that multiple-choice performance does not reliably reflect generative architectural competence.

\FloatBarrier

\begin{figure*}[t]
  \centerline{\includegraphics[width=\textwidth]{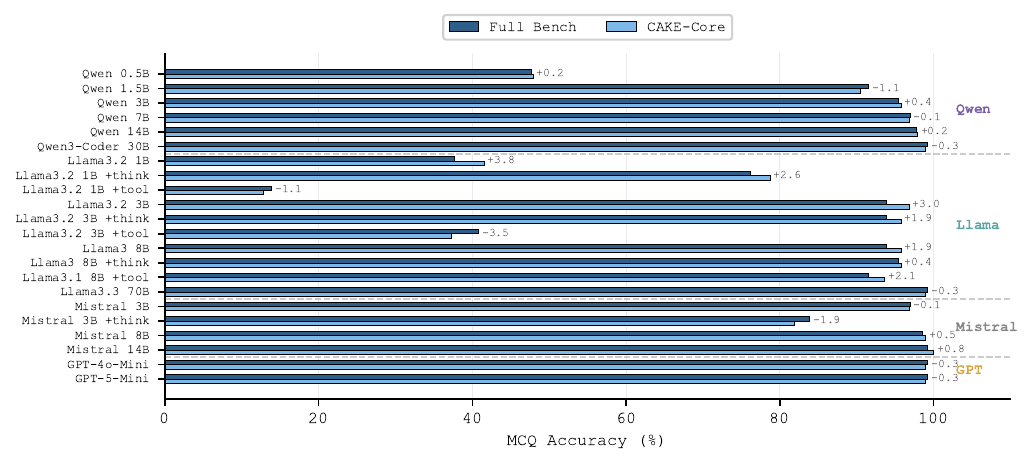}}
  \caption{Full Bench vs.\ CAKE-Core MCQ accuracy for all 22 configurations. Darker bars show Full Bench; lighter bars show CAKE-Core. Delta values at bar ends indicate the accuracy change after quality filtering. Ranking order is largely preserved across model families.}
  \label{fig:core-comparison}
\end{figure*}

\section{Discussion}
\label{sec:discussion}

\subsection{Scaling and Format Effects}

This study provides a benchmark for cloud-native software architecture knowledge, addressing the current lack of a dedicated evaluation resource in this domain. The results indicate that generative evaluation provides a more informative view of architectural understanding than multiple-choice evaluation alone. Implement-level scores capture procedural knowledge with Mistral~14B reaching 4.54 out of max. 5 at the implement level and GPT-5-Mini reaching 4.50 out of max. 5. Training data composition continues to matter. This can be seen with Mistral~3B outperforming Qwen~7B on free-response questions despite having fewer parameters.

\begin{figure}[t]
  \centerline{\includegraphics[width=\columnwidth]{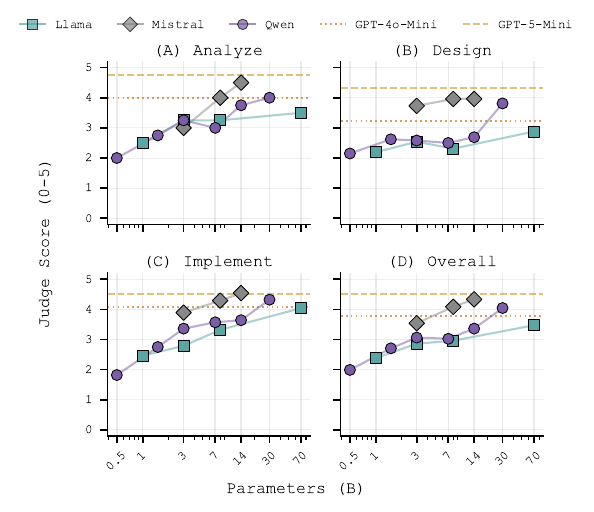}}
  \caption{Free-response judge scores (0--5) vs.\ model parameters across cognitive levels. (A) Analyze, (B) Design, (C) Implement, and (D) Overall. Scores scale consistently with size for all families; GPT model baselines shown as dashed lines. Unlike MCQ (which saturates above 3B), free-response continues differentiating across the full parameter range.}
  \label{fig:freeresponse}
\end{figure}

\begin{figure*}[t]
  \centerline{\includegraphics[width=\textwidth]{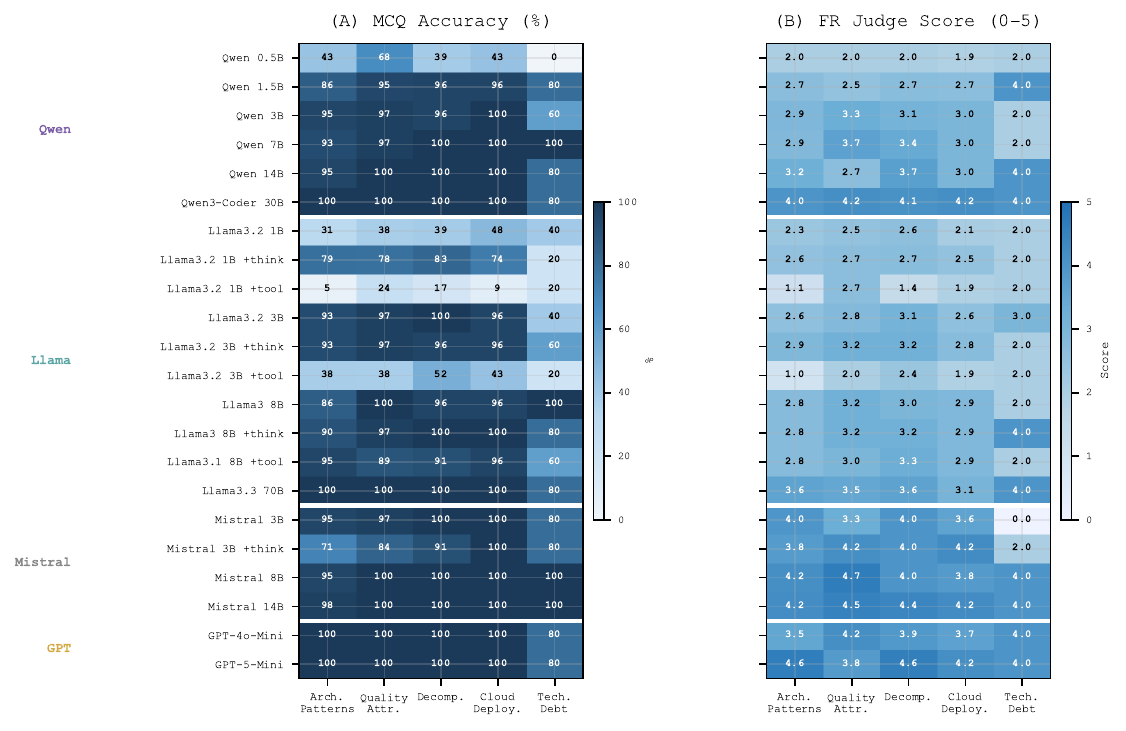}}
  \caption{Per-topic performance across all 22 configurations.
(A) MCQ accuracy (\%) aggregated across recall, analyze, and design levels. (B) free-response judge scores (0--5) aggregated across analyze, design, and implement levels. Color intensity indicates performance; white separators mark family boundaries.}
  \label{fig:topic-performance}
\end{figure*}

At the topic level (Fig.~\ref{fig:topic-performance}), the models show systematic strengths and weaknesses that fade with scale. On recall-level MCQ, Qwen~0.5B scores 72.7\% on quality attributes but only 35.0\% on architectural patterns and 0\% on technical debt. This gap suggests that sub-billion-parameter models retain basic quality vocabulary (e.g., availability, scalability) yet lack coherent knowledge of complex patterns such as CQRS, saga, or sidecar. Above 3B parameters all five topics converge to $\geq$90\%, and this convergence holds at the design level too. One exception stands out, which is cloud deployment at the design level, where Qwen~0.5B scores 0\% but every 3B+ model hits 100\%, pointing to a sharp capability threshold for deployment-related reasoning.

Conviction rates mirror this size dependency. At recall level, Qwen~0.5B gives unanimous answers on just 56\% of questions (28/50), while Qwen~7B reaches 96\% (48/50) and GPT-5-Mini hits 100\%. This rapid jump shows that answer stability emerges alongside accuracy gains, reinforcing conviction as a useful indicator of model confidence rather than mere response consistency.

The conviction metric also offers a practical confidence signal for practitioners. Unanimous answers (conviction~=~1.0) hit 89.5\% accuracy, while split-majority answers (conviction~=~2/3) fall to 55.0\%. That 34.5 percentage-point gap suggests conviction could work as a filter in production: an LLM assistant could flag low-conviction architectural suggestions for human review.

To check how question quality affects results, we built CAKE-Core: a subset of the 188 evaluated questions where every expert rated clarity $\geq$~4, mean correctness $\geq$~4.0, and no flags. Of these, 116 pass (94 MCQ, 22 free-response), distributed as 41 recall, 40 analyze, 19 design, and 16 implement. Fig.~\ref{fig:core-comparison} compares Full Bench and CAKE-Core MCQ accuracy across the 94 core MCQ items for all 22 configurations. Accuracy differences are negligible: Qwen~0.5B shifts from 47.7\% to 47.9\% ($+$0.2\,pp), Qwen~14B from 97.7\% to 97.9\% ($+$0.2\,pp), and GPT-5-Mini from 99.2\% to 98.9\% ($-$0.3\,pp). The near-zero deltas indicate that the quality filter neither inflates nor deflates scores, confirming that the full benchmark is well-calibrated and that CAKE-Core preserves the original ranking order.

\begin{figure}[t]
  \centerline{\includegraphics[width=\columnwidth]{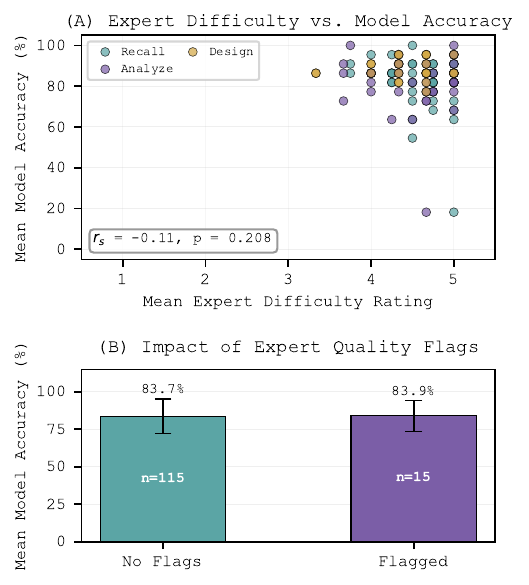}}
  \caption{Expert--model alignment analysis. (A)~Expert difficulty ratings show no significant correlation with mean model accuracy across 22 configurations. (B)~Questions flagged for ambiguity or typos show no accuracy difference from unflagged questions.}
  \label{fig:expert-alignment}
\end{figure}

We also looked at expert--model alignment (Fig.~\ref{fig:expert-alignment}). Panel~(A) shows that per-question expert difficulty ratings don't predict model accuracy (Spearman $r_s = -0.11$, $p = 0.208$). Human and model difficulty perceptions diverge: questions experts call hard are not systematically harder for LLMs, and vice versa. Panel~(B) asks whether expert quality flags (ambiguity or typo) affect model performance. Flagged questions ($n = 15$) and unflagged questions ($n = 115$) yield nearly identical mean model accuracy (83.9\% vs.\ 83.7\%), suggesting that flagged questions aren't inherently harder for models but instead reflect surface-level issues human annotators notice yet LLMs bypass through pattern-matching. Even so, expert correctness ratings remain a meaningful quality signal, since questions unanimously rated 5.0 for correctness produce the highest mean model accuracy (85.3\%) and conviction (75.8\%), compared to 67.0\% accuracy for questions rated below~4.0. This finding validates the role of expert annotation in benchmark curation, especially for spotting questions where the intended correct answer may itself be debatable.

\subsection{Augmentation}

The \texttt{+think} and \texttt{+tool} results have direct practical implications. Reasoning augmentation helps free-response consistently but can destabilize MCQ performance near the ceiling. The MCQ drop under +think appears to come from overthinking. One possible explanation is that extended reasoning chains lead to models incorrectly revising correct answers. Tool augmentation demands a minimum model capacity ($\approx$8B parameters), below that threshold, models produce malformed tool invocations or get stuck in repeated search loops that exceed the iteration limit without adding useful information. From a practitioner's standpoint, a Mistral~3B base model (1.3\,s/question, 3.54 FR overall) may beat a Llama3.2~3B +think variant (4.9\,s/question, 3.01 FR overall) when balancing quality against throughput. For the highest free-response quality, Mistral~14B base (2.4\,s, 4.33 FR overall) offers the best local-model trade-off, approaching GPT-5-Mini (4.52) at a fraction of the latency.

\subsection{Implications and Limitations}\label{sec:limitations}

Models are reliable for MCQ-style knowledge tasks: base models with 3B+ parameters answer recall, analysis, and design questions correctly over 90\% of the time. Free-response evaluation reveals a capability gap that MCQ accuracy alone does not reflect~\cite{b24}. These results address three distinct audiences.

\textit{For practitioners}, model selection should hinge on generative evaluation rather than multiple-choice accuracy~\cite{b8}. The conviction metric supplies an added signal, suggestions produced with unanimous conviction (89.5\% accuracy) can be trusted more than split-majority responses (55.0\%).

\textit{For educators}, the cognitive-level breakdown maps straight to Bloom's taxonomy~\cite{b13}, helping to point out which competencies can be handed off to LLMs. Recall and analysis tasks are well within reach of even small models, while design and implementation require larger models or human oversight.

\textit{For tool builders}, free-response quality suggests LLMs work as effective first-draft generators for architectural artifacts~\cite{b12}, as long as a human architect reviews the output. The wide variance in implement-level scores (1.36--4.54) means implementation assistance should come with confidence-aware guardrails.

The study reveals six key limitations.
\\\textit{1) Scope:} CAKE targets cloud-native architecture and results may not transfer to classical patterns.
\\\textit{2) MCQ distractor quality:} Expert review detected elaboration bias in some distractors. Option shuffling mitigates positional bias and CAKE-Core addresses this with strict quality filtering.
\\\textit{3) Excluded implement MCQs:} 12 implement-level MCQ items were dropped due to a formatting defect, since implementation knowledge is better assessed through free-response questions, the final MCQ set comprised 130 items.
\\\textit{4) Judge model:} Free-response relies on a single primary judge (DeepSeek-R1:32B), partially mitigated by validation with Gemini 2.5 Pro.
\\\textit{5) Inter-rater reliability:} Ordinal Krippendorff's $\alpha$ values are near zero, a known artifact when distributions are heavily skewed~\cite{b16}. While 91.3\% within-one-point agreement supports practical consensus, future iterations should consider wider scales.
\\\textit{6) Correct answer patterns:} Questions tend to have the longest option as the correct answer, which could allow heuristic exploitation; future iterations should control distractor length.

\section{Conclusion}
\label{sec:conclusion}
CAKE is the first benchmark built for cloud-native software architecture knowledge across cognitive levels. Our evaluation of 22 model configurations surfaces four findings. First, MCQ accuracy reaches near-ceiling for models with 3B+ parameters. Second, free-response scores scale steadily across all three cognitive levels, including implement (best: 4.54/5). Third, the two evaluation formats expose complementary facets of knowledge, as MCQ saturates while free-response keeps differentiating models across the full parameter range. Finally augmentation effects depend on size, with reasoning enhancement lifting free-response quality while tool augmentation leads to degradation below $\approx$8B parameters.

These findings extend beyond software architecture. The gap between MCQ and free-response results indicates that benchmarks relying solely on multiple-choice questions may overestimate model capabilities, particularly in domains demanding procedural and design knowledge. Moreover, the conviction metric demonstrates that multi-run evaluation provides actionable confidence signals with no additional cost beyond increased inference time.

Future work will address the coverage gap by incorporating correctly formatted implementation-level MCQs, expand the scope beyond cloud-native architectures to include classical software architecture patterns, investigate optimal augmentation strategies across different model sizes, and extend the benchmark to multilingual settings.

\section*{Data Availability}
The CAKE dataset is publicly available at \\\url{https://github.com/timadam03/CAKE-benchmark}.

\balance

\end{document}